\documentclass[useAMS,usenatbib]{mn2e}
\usepackage{epsfig}
\usepackage{amsmath, amssymb,bm}

\title[Evolution of the Snow Line]{On the evolution of the snow line
  in protoplanetary discs II: Analytic Approximations}

\author[R. G. Martin \& M. Livio]{Rebecca G. Martin$^1$\thanks{E-mail:
  rebecca.martin@jila.colorado.edu} and Mario Livio$^2$\\ $^1$NASA Sagan Fellow, JILA,
University of Colorado, Boulder, CO 80309, USA\\ $^2$Space Telescope
Science Institute, 3700 San Martin Drive, Baltimore, MD 21218, USA
\\ }

\begin{document}

\date{}

\pagerange{\pageref{firstpage}--\pageref{lastpage}} 
\pubyear{2013}
\maketitle

\label{firstpage}

\begin{abstract}
We examine the evolution of the snow line in a protoplanetary disc
that contains a dead zone (a region of zero or low turbulence). The
snow line is within a self-gravitating part of the dead zone, and we
obtain a fully analytic solution for its radius. Our formula could
prove useful for future observational attempts to characterise the
demographics of planets outside the snow line. External sources such
as comic rays or X-rays from the central star can ionise the disc
surface layers and allow the magneto-rotational instability to drive
turbulence there. We show that provided that the surface density in
this layer is less than about $50\,\rm g\,cm^{-2}$, the dead zone
solution exists, after an initial outbursting phase, until the disc is
dispersed by photoevaporation. We demonstrate that the snow line
radius is significantly larger than that predicted by a fully
turbulent disc model, and that in our own solar system it remains
outside of the orbital radius of the Earth. Thus, the inclusion of a
dead zone into a protoplanetary disc model explains how our Earth
formed with very little water.
\end{abstract}


\section{Introduction}

The snow line radius in a protoplanetary disc is the radius beyond
which ice forms, and it occurs at a temperature of around $T_{\rm
  snow}=170\,\rm K$ \citep{lecar06}. It is thought to have very
important consequences for planet formation because the solid mass
density outside of the snow line is much larger due to the
condensation of water.  The current location of the snow line in our
solar system is within the asteroid belt, at $2.7\,\rm AU$. The outer
asteroids are icy C-class objects \citep[e.g.][]{abe00,morbidelli00}
whereas the inner asteroid belt objects contain little water. Thus,
the snow line is thought to have been at its current location since
the time of planetesimal formation within the solar nebula.

Most models of the evolution of the snow line location assume a fully
turbulent steady-state accretion disc including viscous and stellar
heating
\citep[e.g.][]{sasselov00,kennedy06,garaud07,kennedy08,oka11}. Turbulence
is thought to be driven by the magneto-rotational instability (MRI)
\citep{balbus91}, and it transports angular momentum outwards allowing
accretion on to the central star.  However, these models all find that
during the protoplanetary disc evolution the snow line moves inside
the orbital radius of the Earth
\citep[e.g.][]{davis05,garaud07,oka11}. Thus, it is hard to form the
relatively water-devoid Earth in such models \citep[see introduction
  of][for more details]{martin12}.

However, it is now widely accepted that protoplanetary discs contain a
``dead zone'', a region of zero or low turbulence
\citep{gammie96,gammie98}.  \cite{martin12} found that in a
time-dependent disc with a dead zone the radial location of the snow
line is quite different from that in the fully turbulent disc model.
The inner parts of a disc with a dead zone are not in steady state
because material is accumulating in the dead zone. However, the snow
line occurs farther out in the dead zone, where sufficient material
has accumulated that the dead zone is self gravitating. The self
gravity drives a second type of turbulence
\citep[e.g.][]{paczynski78,lodato04}, thus around the snow line
temperature, the disc may be in a self-gravitating locally steady
state \citep{martin13}.

In this work we first revisit the fully turbulent disc model in order
to find an analytic approximation for the snow line radius in that
model. We then develop analytic approximations for the snow line
radius in a model with a dead zone and consider disc parameters for
which this solution is relevant.

\section{The Protoplanetary Disc Model}

Material at radius $R$ in an accretion disc orbits the central star of
mass $M$ at Keplerian velocity, $\Omega=\sqrt{GM/R^3}$. Viscosity
(when it is significant) is generated by turbulence. If the disc is
fully ionised, the gas is well coupled to the magnetic field and
turbulence is driven by the MRI. We consider steady-state solutions of
this kind in Section~\ref{turb}.

\begin{figure}
\includegraphics[width=8.4cm]{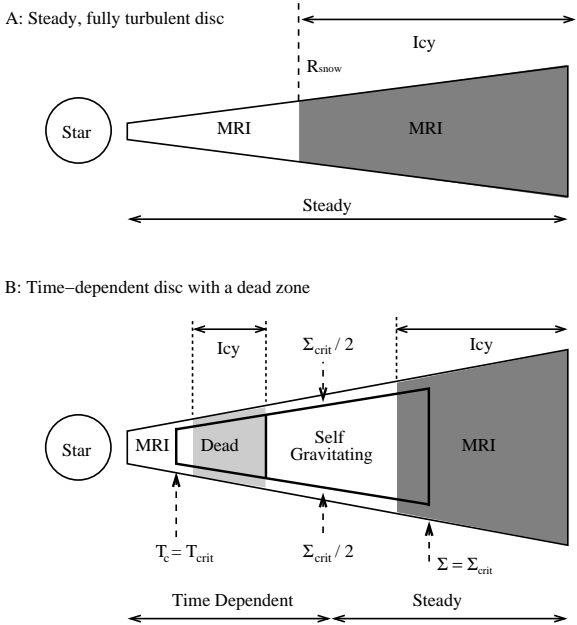}
\caption{Sketches of the disc structure for a steady state fully
  turbulent disc (upper sketch) and for a time-dependent disc with a
  dead zone (lower sketch). For the disc with a dead zone, the outer
  parts of the disc are in steady state but the dead zone causes a
  time-dependent pile up of material in the inner parts. The dead zone
  extends from the radius in the disc where the midplane temperature
  drops below $T_{\rm c}=T_{\rm crit}$ up to the radius where the
  surface density drops below $\Sigma=\Sigma_{\rm crit}$. The shaded
  regions show where the midplane temperature of the disc is below the
  snow line temperature, $T_{\rm c}<T_{\rm snow}$. The outermost snow
  line is at the inner boundary of the darker region and is within the
  self-gravitating part of the dead zone.}
\label{fig}
\end{figure}

However, if the disc is not sufficiently ionised a dead zone may be
present at the mid plane where the MRI does not operate. This causes a
pile up of material in the inner regions. With sufficient
accumulation, the outer parts of the dead zone become self-gravitating
and a second type of turbulence is driven. The snow line in this case
occurs in a steady self gravitating part of the disc \citep{martin12},
and we consider solutions to this configuration in
Section~\ref{dz}. Fig.~\ref{fig} shows a schematic representation of
the two disc solutions.

As the infall accretion rate drops, the dead zone may be accreted at
some point during the evolution, and the disc can transition between
these two solutions. We consider this possibility in
Section~\ref{transition}. Finally in Sections~\ref{irr} and~\ref{phot}
we discuss the effects of irradiation from the central star and
photoevaporation at the end of the disc lifetime, respectively.

\subsection{Fully MRI Turbulent Disc}
\label{turb}

A fully turbulent MRI active disc has viscosity that may be
parameterised with
\begin{equation}
\nu=\alpha_{\rm m}\frac{c_{\rm s}^2}{\Omega},
\end{equation}
where $c_{\rm s}=10^5\, \sqrt{T_{\rm c}/100\,\rm K} \,\rm cm\,s^{-1}$
is the sound speed, $T_{\rm c}$ is the midplane temperature and
$\alpha_{\rm m}$ is the \cite{shakura73} viscosity parameter. The disc
has a steady state surface density
\begin{equation}
\Sigma =\frac{\dot M}{3\pi\nu}
\end{equation}
\citep{pringle81}, where $\dot M$ is the constant infall accretion
rate. In a steady state the accretion rate through all radii is
constant.  The steady surface temperature is given by
\begin{equation}
\sigma T_{\rm e}^4=\frac{9}{8}\frac{\dot M}{3\pi}\Omega^2
\end{equation}
\citep[e.g.][]{cannizzo93,pringle86}.  The midplane temperature is found with
\begin{equation}
 T_{\rm c}^4=\tau T_{\rm e}^4,
\label{t}
\end{equation}
where the optical depth is
\begin{equation}
\tau =\frac{3}{8}\kappa \frac{\Sigma}{2}
\label{tau}
\end{equation}
and the opacity is
\begin{equation}
\kappa=a T_{\rm c}^b.
\label{kappa}
\end{equation}
At temperatures close to the snow line we take $a=3$ and $b=-0.01$
\citep{bell94,bell97}. 

\begin{figure}
\includegraphics[width=8.4cm]{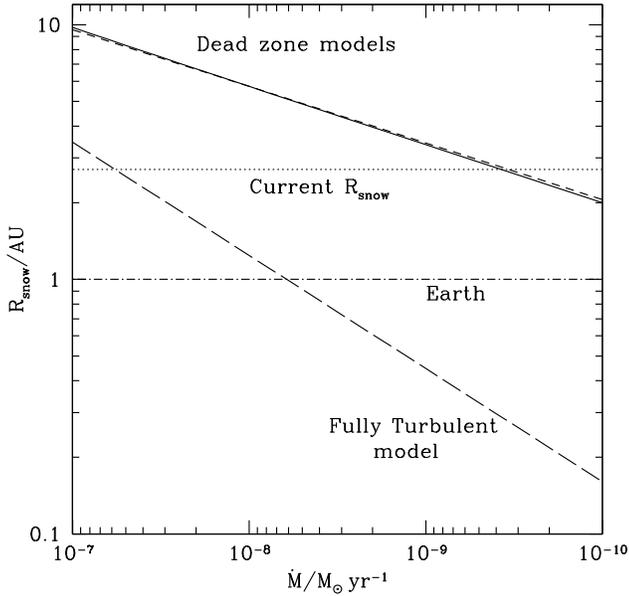}
\caption{The snow line radius, $R_{\rm snow}$, as a function of the
  accretion rate through the disc, $\dot M$. The solid line shows the
  full dead zone solution given in equation~(\ref{full}) and the short
  dashed line shows the approximation given in
  equation~(\ref{app}). The long dashed line shows the fully MRI
  turbulent disc snow line given in equation~(\ref{rturb}). The dotted
  line shows the current position of the snow line in the solar system
  at $R=2.7\,\rm AU$. The dot-dashed line shows the orbital radius of
  the Earth.}
\label{data3}
\end{figure}

We solve the equation $T_{\rm c}=T_{\rm snow}$ and find for the snow
line radius
\begin{align}
R_{\rm snow}   =  & \,\,1.24\,  \left(\frac{\alpha_{\rm m}}{0.01}\right)^{-\frac{2}{9}} \left(\frac{M}{\rm M_\odot}\right)^{\frac{1}{3}} \left(\frac{\dot M}{10^{-8}\,\rm M_\odot\, yr^{-1}}\right)^{\frac{4}{9}} \cr & \times \left(\frac{T_{\rm snow}}{170\,\rm K}\right)^{-\frac{167}{150}} \,\rm AU.
\label{rturb}
\end{align}
This radius as a function of $\dot M$ is shown in Fig.~\ref{data3} by
the long dashed line, for $\alpha_{\rm m}=0.01$, $M=1\,\rm M_\odot$
and $T_{\rm snow}=170\,\rm K$. This functional form is in good
agreement with previous snow line models \citep[e.g.][]{garaud07,
  min11,oka11}.

\subsection{A Disc with a Dead Zone}
\label{dz}

\begin{figure}
\includegraphics[width=8.4cm]{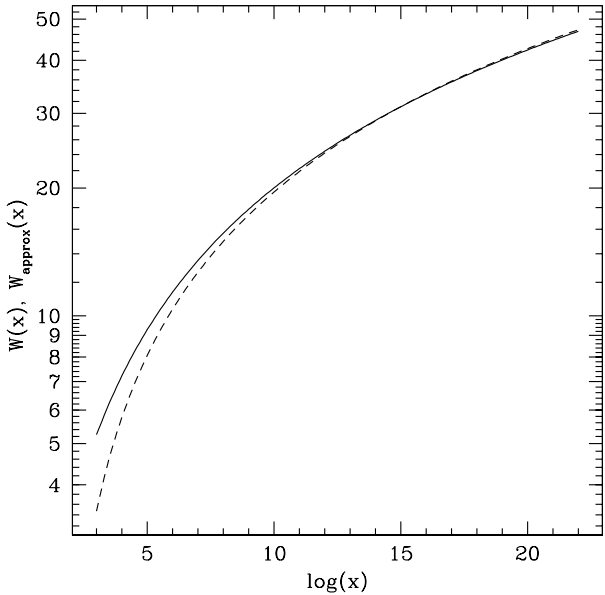}
\caption{The Lambert $W(x)$ function (solid line) and the approximation given in
  equation~(\ref{approx}) (dashed line).}
\label{data1}
\end{figure}

In \cite{martin12} we found that the evolution of the snow line is
significantly altered when a dead zone is introduced in the disc
model. When the infall accretion rate is very high, the disc can be
gravo-magneto unstable and large outbursts in the accretion rate are a
result of transitions from gravitationally produced to magnetically
produced turbulence \citep[][]{armitage01,
  zhu09,martin11,martin13}. The behaviour in this case is extremely
time-dependent. However, once the accretion rate drops below about
$10^{-7}\,\rm M_\odot\,yr^{-1}$, the inflow is insufficient for
outbursts, but the dead zone may remain in the inner parts of the disc
and a steady state is achieved further out in the disc (see
Fig.~\ref{fig}).

The outer parts of the dead zone become self gravitating when the
\cite{toomre64} parameter, $Q=c_{\rm s}\Omega/\pi G\Sigma$, is below
its critical value of $Q_{\rm crit}=2$. This drives a second type of
turbulence with viscosity
\begin{equation}
\nu_{\rm g}=\alpha_{\rm g} \frac{c_{\rm s}^2}{\Omega},
\end{equation}
where we take
\begin{equation}
\alpha_{\rm g}=\alpha_{\rm m} \exp\left(-Q^4\right)
\end{equation}
\citep[e.g.][]{zhu10b}. The precise form of such a viscosity is not
important provided that it is a strongly decreasing function of $Q$
\citep{zhu10a,zhu10b,martin13}. Note that we have taken a different
functional form from that used in \cite{martin12} because with this
prescription we can find analytical solutions more easily.

In a disc with a dead zone in the inner regions, there may be an inner
icy region within the dead zone \citep{martin12}. However, it is not
clear yet whether water is present in such a region and this is still
to be confirmed with detailed simulations including the water
distribution. The outermost snow line occurs in a part of the dead
zone that is self-gravitating. There may be MRI active surface layers
with surface density $\Sigma_{\rm crit}$ \citep[that are ionised by
  external sources such as cosmic rays or X-rays from the central
  star, e.g.][]{glassgold04} on top of the self-gravitating dead zone
(see Fig.~\ref{fig}).  Here we take the limit $\Sigma >>\Sigma_{\rm
  crit}$, and we approximate $\Sigma_{\rm crit}=0$, which allows us to
obtain analytic solutions. Formally, this assumption means that the
dead zone always extends all the way out to the outermost snow line.
The numerical models in \cite{martin12}, with $\Sigma_{\rm crit}>0$,
found this to be the case. We check the validity of the assumption in
the next Section.

With the definition of the Toomre parameter, the surface density of the
self gravitating part of the disc is
\begin{equation}
\Sigma = \frac{c_{\rm s}\Omega}{\pi G Q}.
\label{sig}
\end{equation}
For a steady state disc the accretion rate is
\begin{equation}
\dot M  = 3 \pi \nu_{\rm g}\Sigma .
\end{equation}
Because both $\Sigma$ and $\nu_{\rm g}$ depend on $Q$, this
relates the Toomre parameter to the accretion rate
\begin{equation}
 \dot M = \left(\frac{3 \, c_{\rm s}^3\alpha_{\rm m}}{G}\right) \frac{\exp\left(-Q^4\right) }{Q},
\label{qq}
\end{equation}
where the term in brackets is constant for a fixed snow line
temperature. Because of the sensitive dependence on $Q$, for a
reasonable range of accretion rates, $Q$ is approximately
constant. This was also seen in the numerical simulations of
\cite{martin12}.  Scaling the variables to $T_{\rm snow}'=T_{\rm
  snow}/170\,\rm K$, $\alpha_{\rm m}'=\alpha_{\rm m}/0.01$, $M'=M/\rm
M_\odot$ and $R'=R/\rm AU$, we solve equation~(\ref{qq}) to find the
scaled Toomre parameter, $Q'=Q/Q_{\rm crit}$. For a steady disc we
find
\begin{equation}
Q'=0.81\,\left[ \frac{W(x)}{W(x_0)}\right]^\frac{1}{4},
\label{q}
\end{equation}
where we define
\begin{equation}
x=2.45\times 10^{13} \frac{\alpha_{\rm m}'^4 T_{\rm snow}'^6}{\dot M'^4}
\end{equation}
and $x_0=2.45\times 10^{13}$. $W$ is the Lambert function, defined by
the equation
\begin{equation}
x=W(x)\exp[W(x)].
\end{equation}
For the approximate range $10^{9}<x<10^{21}$ the Lambert function can be
approximated by
\begin{equation}
W_{\rm approx}(x)\approx  \log(x) -3.43.
\label{approx}
\end{equation}
This is appropriate for accretion rates in the range $10^{-7}<\dot
M/{\rm M_\odot\, yr^{-1}}<10^{-10}$.  Fig.~\ref{data1} shows the
Lambert function and the approximation in
equation~(\ref{approx}). Note that $Q'$ is independent of radius in
the disc.

At the snow line, the midplane temperature of the disc is $T_{\rm
  c}=T_{\rm snow}$.  The midplane temperature is related to the disc
surface temperature through equations~(\ref{t})--(\ref{kappa}).
Because we are interested in the snow line temperature of around
$T_{\rm snow}=170\,\rm K$ we again take $a=3$ and $b=-0.01$.  We solve
equation~(\ref{t}) to find the surface temperature as a function of
radius. The steady energy equation is
\begin{equation}
\sigma T_{\rm e}^4=\frac{9}{8}\nu_{\rm g} \Sigma \Omega^2,
\end{equation}
which gives for the snow line
\begin{align}
R_{\rm snow}\,\,  =\,\, &  5.73\, \alpha_{\rm m}'^{\frac{2}{9}} T_{\rm snow}'^{-\frac{67}{150}} M'^{\frac{1}{3}}   \left[\frac{W(x)}{W(x_0)}\right]^{-\frac{1}{9}}   \cr &\times  \exp \left\{-\frac{5}{90}\left[W(x)-W(x_0)\right] \right\}   \,\rm AU.
\label{full}
\end{align}
With the approximation to the Lambert function given in
equation~(\ref{approx}) we find the simple approximation
\begin{equation}
R_{\rm snow} \approx 5.73 \,\left(\frac{M}{\rm M_\odot}\right)^{\frac{1}{3}}\left(\frac{ \dot M}{10^{-8}\,\rm M_\odot\, yr^{-1}}\right)^{\frac{2}{9}}\left(\frac{T_{\rm snow}}{170\,\rm K}\right)^{-\frac{4}{5}} \,\rm AU.
\label{app}
\end{equation}
The two solutions corresponding to equations~(\ref{full})
and~(\ref{app}) are shown in Fig.~\ref{data3} for $M=1\,\rm M_\odot$
and $T_{\rm snow}=170\,\rm K$ (by the solid and short-dashed lines,
respectively). As we can see, the agreement is excellent.  Note that
there is no $\alpha_{\rm m}$ dependence for this dead zone disc
solution.  The dead zone solution is also in good agreement with the
numerical snow line evolution model presented in \cite{martin12}. For
example, in the simulation, the infall accretion rate is $10^{-8}\,\rm
M_\odot\, yr^{-1}$ at a time of $7.6\times 10^5\,\rm yr$ and at this
time the snow line radius is $5.83\,\rm AU$ (see their Fig.~1).

\subsection{Transition Accretion Rate}
\label{transition}

\begin{figure}
\includegraphics[width=8.4cm]{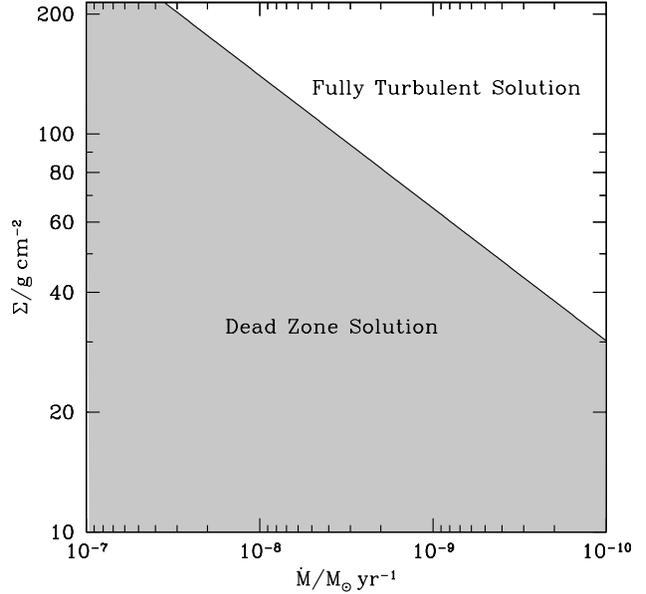}
\caption{The surface density in the disc at the radius $R=R_{\rm
    crit}$ (which is where $T_{\rm c}=T_{\rm crit}$) as a function of
  the accretion rate through the disc. If $\Sigma_{\rm crit}$ is
  smaller, the dead zone solution exists (shaded region). If
  $\Sigma_{\rm crit}$ is larger, the disc is fully turbulent (unshaded
  region).}
\label{turbfig}
\end{figure}

As the accretion rate through the disc drops, it is possible for the
dead zone to eventually be accreted on to the star. 
This situation may arise if the external ionisation sources are strong
enough to ionise the entire disc, as it becomes depleted.  If this
happens, then the disc will transition from the dead zone solution
described in Section~\ref{dz} to the fully turbulent solution
described in Section~\ref{turb}.

The inner parts of the disc are thermally ionised and the MRI operates
out to the radius in the disc where the temperature drops below some
critical value, $T_{\rm crit}$ (see Fig.~\ref{fig}). This is thought
to be around $800\,\rm K$ \citep{umebayashi83}. Farther out in the
disc, the external sources (cosmic rays or X-rays from the central
star) ionise only the surface layers so that the MRI operates in a
layer with surface density $\Sigma_{\rm crit}$.  The transition from
the dead zone solution to the fully turbulent solution will occur when
the surface density in the fully turbulent disc (at the radius where
$T_{\rm c}=T_{\rm crit}$) drops below $\Sigma_{\rm crit}$. Once this
occurs, the whole disc becomes fully MRI active.

The value of $\Sigma_{\rm crit}$ is not well determined. If cosmic
rays are the dominant source of ionisation, $\Sigma_{\rm crit}\approx
200\,\rm g\,cm^{-2}$ \citep{gammie96,fromang02}. However, in the
absence of cosmic rays, X-rays may be the dominant source and in this
case the active layer surface density is much smaller
\citep{matsumura03}.  The ionisation is further suppressed with the
inclusion of effects such as ambipolar diffusion and the presence of
dust and polycyclic aromatic hydrocarbons
\citep{bai09,perez11,simon12,dzyurkevich13}.  In order to explain the
observed T Tauri accretion rates a value $\Sigma_{\rm crit}>10\,\rm
g\,cm^{-2}$ is required \citep{perez11}.

The radius where $T_{\rm c}=T_{\rm crit}$ was found by \cite{martin13}
to be
\begin{equation}
R_{\rm crit} = 0.24\,\alpha_{\rm m}'^{-\frac{2}{9}}M'^{\frac{1}{3}} \dot M'^{\frac{4}{9}} T_{\rm crit}'^{-\frac{14}{15}}\,\rm AU,
\end{equation}
where $T_{\rm crit}'=T_{\rm crit}/800\,\rm K$. They solved the fully
turbulent disc equations described in Section~\ref{turb} but with an
opacity law of $\kappa=0.02\,T_{\rm c}^{0.8}$ \citep{bell94,bell97},
because the temperatures here are higher. In Fig.~\ref{turbfig} we
show the surface density of the fully MRI turbulent disc as a function
of the accretion rate at $R_{\rm crit}$ with $\alpha_{\rm m}=0.01$,
$M=1\,\rm M_\odot$ and $T_{\rm crit}=800\,\rm K$. If $\Sigma_{\rm
  crit}$ lies within the shaded region, the dead zone solution exists,
whereas if $\Sigma_{\rm crit}$ lies in the unshaded region, the whole
disc is MRI turbulent. If, for example, $\Sigma_{\rm crit}=200\,\rm
g\,cm^{-2}$, then the dead zone will be accreted once the infall
accretion rate drops to below $2.9\times 10^{-8}\,\rm
M_\odot\,yr^{-1}$ and the disc will transition to the fully turbulent
solution. However, if $\Sigma_{\rm crit}<30\,\rm g\,cm^{-2}$, then the
dead zone solution exists until the accretion rate drops to less than
$10^{-10}\,\rm M_\odot\,yr^{-1}$.

The total surface density at the snow line radius can be approximated
by
\begin{equation}
\Sigma_{\rm snow} \approx 5571\, \dot M'^{-\frac{1}{3}} T_{\rm
  snow}'^{\frac{5}{3}} \,\rm g\,cm^{-2}.
\end{equation}
This is the surface density for the self-gravitating disc solution
given in equation~(\ref{sig}), evaluated at a radius $R=R_{\rm snow}$
and temperature $T_{\rm c}=T_{\rm snow}$ and where $Q$ is related to
the infall accretion rate through equation~(\ref{q}). The dependence
on $\alpha_{\rm m}$ and $M$ in the surface density cancels out.  The
surface density increases with decreasing accretion rate. The surface
density for a high accretion rate of $\dot M=10^{-7}\,\rm
M_\odot\,yr^{-1}$ is $2758\,\rm g\,cm^{-2}$. At the minimum, the
surface density is at least an order of magnitude larger than the
surface density in an active layer that is ionised by external sources
(such as cosmic rays or X-rays). Thus, we were justified in neglecting
the active layers in our self-gravitating dead zone solution in
Section~\ref{dz}.

\subsection{Stellar Irradiation}
\label{irr}

We have not included the effects of irradiation from the central star
in our models. When there is a dead zone present, the irradiation will
have little effect on the solution because the heating is dominated by
that from self-gravitational turbulence. In the fully MRI turbulent
solution, once the accretion rate becomes very low, $\lesssim
10^{-9.5}\,\rm M_\odot\,yr^{-1}$, the irradiation becomes dominant and
the snow line radius is approximately constant (with accretion rate)
at a value of around $2.2\,\rm AU$ \citep{garaud07}. Similarly,
\cite{oka11} found an increase with decreasing accretion rate for
these low rates because the disc becomes optically thin as the
accretion rate drops. Thus, the effect of including the irradiation
would be to increase the snow line radius slightly for the smallest
accretion rates.

\subsection{Photoevaporation}
\label{phot}

Eventually the disc is thought to be dispersed by photoevaporation by
the ultraviolet radiation from the central star
\citep[e.g.][]{hollenbach94}. The ionising radiation produces a wind
at large radii \citep[typically around $9\,\rm AU$,][]{alexander06b}
where the material becomes unbound. This process begins once the
accretion rate drops to a few $10^{-10}\,\rm M_\odot\,yr^{-1}$, and
the disc is thought to be photoevaporated on a short timescale of
around $10^5\,\rm yr$
\citep[e.g][]{clarke01,alexander06,alexander06b}.

Because our disc models are in steady state, it would be hard to
include such effects. However, qualitatively, photoevaporation would
have a similar effect to that of decreasing the infall accretion rate.
Provided that the dead zone exists until the disc is photoevaporated
(this is true for $\Sigma_{\rm crit}\lesssim 50\,\rm g\,cm^{-2}$, see
Fig.~\ref{turbfig}), the dead zone solution will be appropriate for
the entire disc lifetime (after the initial outburst phase). In this
case, the dead zone model predicts that the snow line in the solar
system reaches its current value of $2.7\,\rm AU$ at an infall
accretion rate of $3.2\times 10^{-10}\,\rm M_\odot\, yr^{-1}$. In
order for the current snow line radius to be explained, the disc must
be photoevaporated around this accretion rate.

\section{Discussion}

The snow line has important implications for planet formation. It is
thought that while gas giant planets form outside of the snow line,
the rocky terrestrial planets form inside the snow line. The location
of an asteroid belt (if one forms at all) may also be a direct
consequence of the snow line. Giant planets form outside of the snow
line thus preventing formation at and around the snow line and leaving
an asteroid belt \citep{martin13b}. Knowing the location of the snow
line may be important for future attempts to characterise the
demographics of planets outside the snow line (such as the proposed
WFIRST).

We have assumed that the surface density that is ionised by external
sources is constant in radius. A more realistic way to find the dead
zone may be with a critical magnetic Reynolds number
\citep[e.g.][]{matsumura03}. However, the accretion rates produced by
such models are too small to account for the observed T Tauri
accretion rates \citep[][]{martin12a,martin12b}.

Shearing box simulations suggest that the turbulence in the MRI active
surface layers may drive a small but non zero hydrodynamic turbulence
in the dead zone \citep[e.g][]{fleming03,turner08,simon11}. Providing
that the turbulence is not strong enough for the disc to reach a
steady state (which is unlikely for realistic disc parameters), the
disc model presented here will not be significantly affected. A
variation of turbulence with height in a snow line model has been
previously considered \citep{kretke07,kretke10}. However, the
turbulence they included in the dead zone was sufficiently high for a
steady state and thus the snow line radius was not significantly
different from that in the fully turbulent model.

There is an inner icy region within the dead zone predicted by our
model. This is within the time dependent region of the disc. The inner
edge of the icy region depends on the irradiation from the central
star whereas the outer edge occurs where the disc first becomes
self-gravitating. As material in the disc builds up, this radius moves
inwards in time. As discussed in \cite{martin12}, this region could,
in principle at least, allow for the formation of hot Jupiters, close
to the star, without the need for significant migration. However, the
evolution of the distribution of water within a disc with a dead zone
should be investigated in future work to fully assess the validity of
this scenario.

\section{Conclusions}

We have found an analytic expression for the radius of the snow line
as a function of the infall accretion rate for a disc with a dead
zone. This solution is appropriate for modelling the entire disc
lifetime after the initial outbursting phase, up to the dispersal by
photoevaporation, provided that the surface density in the MRI active
surface layers that are ionised by cosmic rays or X-rays satisfies
$\Sigma_{\rm crit}\lesssim 50\,\rm g\,cm^{-2}$. Once the infall
accretion rate drops to a few $10^{-10}\,\rm M_\odot\,yr^{-1}$, the
disc is dispersed by photoevaporation leaving the snow line at the
radius where it is observed today, in the solar system at $2.7\,\rm
AU$.

With the dead zone model, unlike in the fully turbulent model, the
snow line does not pass inside the orbital radius of the Earth. Thus,
with this model the formation of our water devoid Earth is possible.

\section*{Acknowledgements}

RGM's support was provided in part under contract with the California
Institute of Technology (Caltech) funded by NASA through the Sagan
Fellowship Program. 


\label{lastpage}
\end{document}